\documentclass[runningheads]{llncs}
\usepackage[utf8]{inputenc}
\usepackage{wrapfig}
\usepackage{lscape}
\usepackage{rotating}
\usepackage{graphicx}
\usepackage{caption}
\usepackage{amsmath, nccmath}
\usepackage{comment}
\usepackage{afterpage}
\usepackage{multirow}
\usepackage{multicol}
\usepackage{cleveref}
\usepackage{rotating}
\usepackage{makecell}
\usepackage{booktabs}
\usepackage{multirow}
\usepackage{lscape}
\usepackage{longtable}
\usepackage{placeins}
\usepackage{adjustbox}
\usepackage{arydshln} 
\usepackage{ragged2e}
\usepackage{lipsum}
\usepackage{mathtools} 
\usepackage{enumitem} 
\usepackage{algorithmic}
\usepackage{xcolor}
\usepackage[ruled,vlined]{algorithm2e}

\usepackage[T1]{fontenc}

\begin{document}

\title{One Word is Enough: Minimal Adversarial Perturbations for Neural Text Ranking}
\titlerunning{Minimal Adversarial Perturbation for Neural Text Ranking}

\author{Tanmay Karmakar\orcidID{0009-0007-1030-0444} \and
Sourav Saha\orcidID{0000-0001-8091-0685} \and
Debapriyo Majumdar\orcidID{0009-0000-8823-9742} \and
Surjyanee~Halder\orcidID{0009-0005-0851-0472}}

\authorrunning{Karmakar et al.}

\institute{Indian Statistical Institute, Kolkata, India\\
\email{tanmayknowsmath@gmail.com, sourav.saha\_r@isical.ac.in, debapriyo@isical.ac.in, surjyaneeh@gmail.com}}

\maketitle              
\begin{abstract}
Neural ranking models (NRMs) achieve strong retrieval effectiveness, yet prior work has shown they are vulnerable to adversarial perturbations. We revisit this robustness question with a minimal, query-aware attack that promotes a target document by inserting or substituting a single, semantically aligned word - the query center. We study heuristic and gradient-guided variants, including a white-box method that identifies influential insertion points. On TREC-DL 2019/2020 with BERT and monoT5 re-rankers, our single-word attacks achieve up to 91\% success while modifying fewer than two tokens per document on average, achieving competitive rank and score boosts with far fewer edits 
under a comparable white-box setup to ensure fair evaluation against PRADA.
 We also introduce new diagnostic metrics to analyze attack sensitivity beyond aggregate success rates. Our analysis reveals a \emph{Goldilocks zone} in which mid-ranked documents are most vulnerable. These findings demonstrate practical risks and motivate future defenses for robust neural ranking. 


\end{abstract}

\section{Introduction}

Neural ranking models (NRMs) based on large Transformer architectures such as
BERT and T5 have achieved state-of-the-art effectiveness across a wide range of
search and ranking tasks. Their ability to capture deep semantic relationships
between queries and documents has made them the dominant paradigm in modern
retrieval pipelines. However, prior work has shown that these models are
vulnerable to small, carefully chosen adversarial text
modifications that can significantly alter ranking outcomes without changing
human-perceived meaning.

Many existing adversarial attacks on NRMs, such as PRADA~\cite{wu2023prada},
rely on multi-word manipulations or surrogate model training, often requiring
substantial text edits or query probing. 
This study revisits the robustness question from a minimalist perspective. 
We demonstrate that inserting or
substituting \emph{a single, query-aware word} can be sufficient to promote a
target document's rank in strong neural rankers. Our approach introduces the
concept of a \emph{query center}---a semantically central term derived from the
user query---and uses it to construct effective, query-guided perturbations.

We present methods based on both heuristic and gradient-guided  variants to identify influential insertion positions. 
Experiments on standard
benchmarks, namely TREC-DL 2019~\cite{craswell2020overviewtrec2019deep} and 2020~\cite{craswell2021overviewtrec2020deep} with two widely used rankers,
BERT-base-mdoc-BM25 and monoT5-base-MSMARCO~\cite{monoT52020}, reveal that these minimal
perturbations can achieve up to 91\% attack success while modifying only one token per document. 
While our gradient-guided approach uses white-box access for practical feasibility, we ensure fairness by evaluating PRADA under the same conditions.
Our method achieves competitive rank and score boosts relative to PRADA
while requiring far fewer edits.

In addition, our analysis reveals a ``Goldilocks zone'' in which mid-ranked
documents are most vulnerable to such minimal attacks. These findings highlight
the fragility of neural ranking models to semantically plausible perturbations
and motivate the development of more robust ranking architectures.


\section{Related Work}

Adversarial robustness of neural language models has been an active area of
study across both text classification and ranking. Early work on textual
attacks focused on character- or word-level perturbations designed to fool
sequence classifiers~\cite{ebrahimi2018hotflip,pruthi2019combating}, later
adapted to information retrieval (IR) and re-ranking tasks.
These studies demonstrated that even minor token-level changes could degrade
performance in neural architectures that rely on subword tokenization.

Wu et~al.~\cite{wu2023prada} introduced PRADA, a black-box framework
for practical adversarial attacks on neural ranking models (NRMs). PRADA trains
a surrogate model using pseudo-relevance feedback and applies gradient-guided
word substitutions to promote target documents. Although highly effective, the
approach requires multiple perturbations per document and repeated model
queries. Subsequent white-box work, such as Wang et~al.~\cite{bert-rankers-brittle},
analyzed the sensitivity of BERT-based rankers using direct gradient access,
showing that perturbing a few words can drastically alter ranking scores. More recent works ~\cite{sigir_24_attack,aaai_25_attack} also focus on achieving very high attack success rate up to almost 100\% but do not emphasize on the minimality of the perturbation. Among other recent works, Bidgeli et~al.~\cite{bigdeli2025empr} proposed a black-box attack strategies via embedding-level perturbations avoiding discrete word edits, and Liu et~al.~\cite{geisler2025reinforce} proposed a  multi-granular adversarial attacks combining word, phrase, and sentence perturbations using reinforcement learning.
Ivgi and Berant~\cite{ivgi-berant-2021-achieving} explored discrete adversarial training
strategies for NRMs, comparing offline and online augmentation regimes. Their
results suggested that online adversarial sampling yields greater robustness, but
at significant computational cost. 

Our work focuses on a \emph{minimal, query-aware} setting where inserting or substituting
a single word -- the \emph{query center} -- is sufficient to affect ranking
outcomes. This work complements existing attack frameworks by exposing the
fragility of NRMs under the smallest possible perturbations while maintaining
semantic coherence.

\section{One-word Adversarial Attack Framework}
\label{sec:adversarial_framework}

We consider the task of promoting a target document within an NRM-based ranked list through minimal textual modification -- specifically, insertion or substitution of a single word.
Given a query $q$ and a ranked list $L$ (reranked by an NRM) over the top-$100$ BM25 candidates, and a target document $d$ in $L$, our goal is to perturb $d$ minimally to obtain $\tilde{d}$ such that the NRM ranks $\tilde{d}$ higher (better) than $d$.
The core idea of our approach is to identify the most influential word in the user
query -- the \emph{query center} -- and insert or substitute it in the target
document at a strategically chosen location, in order to maximize the change in the NRM score while preserving document coherence.

\subsection{The query center}
\label{sec:query_center}
We introduce the notion of the \emph{query center} as the token that is closest to the centroid of the query in a given semantic embedding space. Intuitively, the query center represents a single word that best captures the overall semantic intent of the query.

Let the input query be \( q = (q_1, q_2, \ldots, q_k) \), where each term \(q_i\) is mapped to a vector \( \mathbf{v}(q_i) \) in an embedding space. The centroid of the query is computed as:
\[
    \mathbf{v}(q) = \frac{1}{k} \sum_{i=1}^{k} \mathbf{v}(q_i).
\]
We then identify the token $q_{\text{center}}$ in the embedding vocabulary whose vector is closest to $\mathbf{v}(q)$ under cosine similarity; this token is chosen as the \emph{query center}. Tokens not present in the embedding vocabulary are ignored when computing the centroid.

In our experiments, we instantiate this definition using the publicly released counter-fitted word embeddings\footnote{\url{https://github.com/nmrksic/counter-fitting/tree/master/word\_vectors}} of Mrkšić et al.~\cite{Nicola2016counter_fitted}. These embeddings are derived from pre-trained 300-dimensional GloVe vectors trained on the Common Crawl corpus (840B tokens)~\cite{pennington2014glove} and constructed using lexical semantic constraints from WordNet~\cite{miller1995wordnet} to better capture synonymy and antonymy relations.

The counter-fitted embedding space is used only for semantic operations such as query center selection and similarity computation. All gradient-based computations are performed with respect to the ranker’s own input embeddings.

\subsection{Position of the attack}
\label{sec:attacking_method}
We propose three approaches, namely (1) \texttt{one\_word\_start}, 
(2) \texttt{one\_word\_sim}, and
(3) \texttt{one\_word\_best\_grad}, to determine the position in the document to be aimed for perturbation. The \texttt{one\_word\_start} and \texttt{one\_word\_best\_grad} strategies perform insertions, while \texttt{one\_word\_sim} performs a substitution.

The simplest \texttt{one\_word\_start} variant inserts the query center 
\( q_{\text{center}} \) at the beginning of the target document \( d \).  
This straightforward heuristic is motivated by the positional sensitivity of neural rankers: prior work has shown that BERT-based rankers can be brittle to perturbations at early positions and often place disproportionate weight on the beginning of a passage~\cite{bert-rankers-brittle}. 

The \texttt{one\_word\_sim} strategy substitutes the word most similar to the query center. For each token \( t_i \) in the 
document, we compute its embedding similarity to \( q_{\text{center}} \)
(e.g., cosine similarity in the counter-fitted embedding space) and identify 
the token \( t^* \) that is most similar to \( q_{\text{center}} \) but not 
identical to it. The chosen token \( t^* \) is then replaced with 
\( q_{\text{center}} \), yielding a minimally perturbed document that remains 
semantically natural yet better aligned with the query intent. The first two methods use simple heuristics and require no access to the model internals.

Inspired by PRADA~\cite{wu2023prada}, the \texttt{one\_word\_best\_grad} strategy uses gradient information of the model to determine the most impactful position of the document to attack. 
Given the query $q$, document $d$, a neural ranker $f$, and a ranked list $L$, the pairwise hinge loss is defined by:
\begin{equation} \label{eq:hinge_loss}
    \mathcal{L}_{\text{rank}}(q, d; L) = \sum_{d' \in L \setminus \{d\}} \max(0, \beta - f(q, d) + f(q, d')),
\end{equation}
where \(\beta\) is a margin. The hinge loss measures how strongly the current ranking violates the desired ordering, i.e., how far the target document is from outranking competing documents by a margin.

For each token \(t_i\) in \(d\), we compute the norm of the gradient of the hinge loss with respect to its input embedding: 
\begin{equation} \label{eq:imp_position}
    I(t_i) = \left\| \frac{\partial \mathcal{L}_{\text{rank}}}{\partial \mathbf{e}_{t_i}} \right\|_2^2,
\end{equation}
where \(\mathbf{e}_{t_i}\) denotes the input embedding of token \(t_i\) in the ranker. 
Intuitively, tokens with higher gradient norm exert greater influence on the ranking objective, and perturbing them is more likely to change the rank of the document. 

Hence, we select the top-$k$ positions by the gradient norm and evaluate each candidate by computing the resulting score $f(q,\tilde{d})$ of the document after inserting the query center at each of the top-$k$ positions; the insertion that yields the largest score
increase is chosen by the \texttt{one\_word\_best\_grad} strategy, a pseudocode of which is given in Algorithm \ref{alg:best_grad}. The choice of $k$ balances search cost and effectiveness (we use
\(k=20\) in our experiments).

\begin{algorithm}[t]
\caption{\texttt{one\_word\_best\_grad}}
\label{alg:best_grad}
\KwIn{Query $q$, document $d$, ranked list $L$, ranker $f$, query center $q_{\text{center}}$, candidate size $k$}
\KwOut{Perturbed document $\tilde{d}$}

Compute gradient norm $I(t_i)$ for all tokens $t_i \in d$ using Eq. \ref{eq:imp_position}\;
Select top-$k$ token positions $\{i_1, \dots, i_k\}$ by importance score\;

\For{each $i_j \in \{i_1, \dots, i_k\}$}{
    Insert $q_{\text{center}}$ at position $i_j$ to obtain candidate $\tilde{d}_{i_j}$\;
    Compute score $f(q, \tilde{d}_{i_j})$\;
}

Return the $\tilde{d}_{i_j}$ with the highest score\;
\end{algorithm}

These three methods span a spectrum from black-box heuristics to white-box gradient-guided insertions, all constrained to a single-word perturbation to maximize semantic plausibility while exposing the ranker's fragility. 

\section{Experiments}
\label{sec:experiments}

We evaluate our proposed one-word adversarial attacks on two neural ranking
models, comparing against the PRADA under identical white-box
conditions. 

\subsection{Datasets, models and baseline}

We conducted our study on the MSMARCO v1 passage ranking dataset, which contains approximately 8.8 million passages. We evaluated on two benchmark query sets: TREC DL 2019 and TREC DL 2020, with 43 and 54 queries, respectively. We used two NRMs: a BERT-based model (BERT-base-mdoc-BM25) and a MonoT5-based reranker (monoT5-base-MSMARCO), both of which are strong and widely used neural baselines. In our pipeline, we retrieved the top-$k$ (100) documents using BM25 and reranked them using these NRMs. Our code is publicly available at:
\url{https://github.com/Tanmay-Karmakar/one\_word}.

The original PRADA~\cite{wu2023prada} is a black-box attack that trains a surrogate model to approximate the target NRM and uses its gradients to generate adversarial perturbations. Due to computational constraints, we instead use PRADA directly in a white-box setting, ensuring a fair comparison with our gradient-guided variants by giving all methods identical parameter access.

\subsection{Evaluation metrics}
An attack is said to be successful if the rank of the perturbed document $\tilde{d}$ is higher when it replaces the original document $d$ and the list $L$ is re-ranked by the NRM. We report \textit{Success Rate} (SR), the percentage of attacks that promote a document’s rank, and 
 \textit{Similarity Score} (SS): the cosine similarity (a score in $[0,1]$) between the original document embedding and the perturbed document embedding. The embeddings are computed using the Universal Sentence Encoder\footnote{\url{https://tfhub.dev/google/universal-sentence-encoder/4}}.
 An adversarial attack aims to achieve high SR while maintaining high SS. We also report \textit{Rank Boost} (RB) and \textit{Score Boost} (SB), which measure the average improvement in rank position and NRM score, respectively.

In addition, we introduce two fine-grained metrics to capture the structure of ranking changes. \textit{Perturbation Percentage} (PP) is the average percentage of tokens modified per document, measuring attack minimality. \textit{Interval Success Rate} (ISR) is the average success rate computed over rank intervals (\([11\text{–}20], [21\text{–}30], \dots, [99\text{–}100]\)), revealing where the model is most fragile; we exclude the top-10 documents as they are unlikely to be further promoted~\cite{wu2023prada}.

\subsection{Results and analysis}

The experimental results, shown in Table~\ref{tab:attack_results}, demonstrate that our best approach, \texttt{one\_word\_best\_grad}, attains up to \textbf{91\%} success rate with only one token change, while maintaining high cosine similarity (\(\approx0.98\)) between the original document \(d\) and the perturbed document \(\tilde{d}\). In contrast, PRADA attains similarity scores below 90\%, indicating substantially larger perturbations. 

\begin{table*}[ht]
\centering
\begin{tabular}{lccccc c ccccc}
\toprule
& \multicolumn{5}{c}{\textbf{TREC DL 2019}} & & \multicolumn{5}{c}{\textbf{TREC DL 2020}} \\
\cmidrule(lr){2-6} \cmidrule(lr){8-12}

\textbf{Attack Method} & \textbf{SR} & \textbf{SS} & \textbf{PP} & \textbf{RB} & \textbf{SB} & &  \textbf{SR} & \textbf{SS} & \textbf{PP} & \textbf{RB} & \textbf{SB} \\
\midrule
& \multicolumn{11}{c}{\textbf{BERT-base-mdoc-BM25}} \\
\midrule
PRADA (Baseline) & 96.69 & 0.87 & 12.30 & 17.84 & 1.98 &  &
97.74 & 0.85 & 13.83 & 20.10 & 2.42 \\
\addlinespace[0.3em]
\texttt{one\_word\_start} & 56.64 & 0.96 & 0.98 & 8.20 & 1.08 &  &
44.18 & 0.98 & 0.81 & 4.20 & 0.53\\
\texttt{one\_word\_sim} & 38.09 & 0.98 & 0.81 & 4.59 & 0.56 & &
30.90 & 0.98 & 0.70 & 2.06 & 0.24 
\\
\texttt{one\_word\_best\_grad} & 90.97 & 0.97 & 1.49 & 12.97 & 1.61 & ~~~ &
87.10 & 0.97 & 1.45 & 9.47 & 1.16 
\\
\midrule
& \multicolumn{11}{c}{\textbf{monoT5-base-MSMARCO}} \\
\midrule
PRADA (Baseline)               & 69.78 & 0.89 & 9.70 & 11.61 & 1.39 & &  59.62 & 0.88 & 3.55 & 7.42 & 0.94 \\
\addlinespace[0.3em]
\texttt{one\_word\_start}      & 57.48 & 0.97 & 0.96 & 9.21 & 1.27 &  & 50.23 & 0.97 & 0.87 & 5.57 & 0.70 \\
\texttt{one\_word\_sim}        & 53.78 & 0.98 & 0.91 & 7.45 & 0.98 & ~ & 45.53 & 0.98 & 0.80 & 4.34 & 0.52 \\
\texttt{one\_word\_best\_grad} & 65.18 & 0.97 & 1.15 & 11.78 & 1.53 & & 59.49 & 0.96 & 1.10 & 7.95 & 1.00 \\
\bottomrule
\end{tabular}
\vspace{0.2cm}
\caption{Attack performance comparison on TREC DL 2019 (left) and TREC DL 2020 (right) for the neural rerankers. SR and PP are in percentages, SS is a fraction in $[0,1]$, while RB and SB are absolute measures.}
\label{tab:attack_results}
\end{table*}

Our best approach also achieves rank and score boosts comparable to PRADA. Even simple heuristics such as \texttt{one\_word\_start} and \texttt{one\_word\_sim} achieve non-negligible success (30--60\%) with a single-word change, while preserving high semantic similarity, highlighting the vulnerability of the NRMs. Across both rankers, rank boost and score boost exhibit a strong positive correlation, confirming that minimal edits can meaningfully shift model confidence.

\begin{figure}[ht]
    \centering
    \includegraphics[width=1\linewidth]{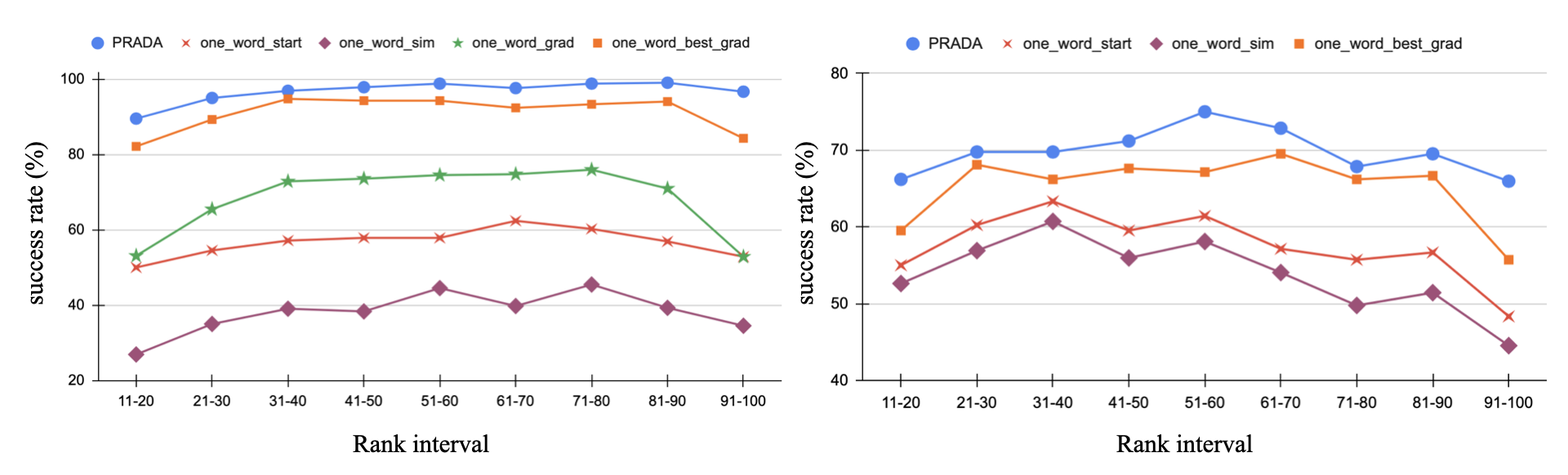}
    \caption{ISR for different attack approaches on BERT-base-mdoc-BM25 (left) and monoT5-base-MSMARCO (right), both using TREC DL 2019 queries.}
    \label{fig:isr_plot}
\end{figure}

Figure~\ref{fig:isr_plot} illustrates the Interval Success Rate, revealing a \textit{Goldilocks zone} in which mid-ranked documents (roughly ranks 40–80) are most susceptible to perturbation. A possible explanation for this is that the top-ranked passages already saturate the relevance score, while bottom-ranked ones are typically too semantically distant to benefit from a single-word insertion. 
Despite occurring away from the top of the ranking, these shifts can still affect multi-stage retrieval pipelines due to aggressive candidate pruning.
 Our ISR analysis thus highlights a vulnerability of neural rankers not captured by aggregate success metrics alone.

\section{Conclusion and Future Work}
\label{sec:conclusion}
We have presented a minimal, query-aware adversarial framework for neural
ranking that perturbs a document with only a single, semantically aligned
word. Despite this simplicity, our attacks substantially alter ranking
outcomes while maintaining over 97\% similarity to the original text,
demonstrating that even moderate success rates (around 50\%) can pose
practical risks in multi-stage neural retrieval pipelines. 
Beyond proposing lightweight attack methods, we introduced two diagnostic metrics—Perturbation Percentage and Interval Success Rate—for deeper, model-agnostic analysis of adversarial behavior, and showed that mid-ranked documents are most vulnerable, highlighting uneven robustness across the ranked list.

In future work, we aim to extend our gradient-based strategy to realistic \emph{black-box} settings and recent LLM-based rankers, and to explore \emph{multiword but still minimal} perturbations to study the trade-off between subtlety and attack success. 
Finally, applying such perturbations to related tasks (e.g., conversational search, question answering, or recommendation) would help assess their broader implications for robust information access systems.


\subsubsection*{\textbf{Acknowledgements.}}
\small We thank the anonymous reviewers for their insightful and constructive feedback, which substantially improved the clarity and quality of this work. Sourav Saha acknowledges the support from the TCS Research Scholar program.

\subsubsection*{\textbf{Disclosure of Interests.}}
\small The authors have no competing interests to declare that are relevant to the content of this article.
\bibliographystyle{splncs04}
\bibliography{references} 

\end{document}